\documentstyle{mn}
\title{The search and investigation of the Large Groups of Quasars}       
\author[]
       { B.V.Komberg $^{1}$, A.V.Kravtsov $^{2}$ and V.N.Lukash $^{1}$\\
 $^{1}$ Astro Space Center, Lebedev Physical Institute, 
  Profsoyuznaya 84/32, 117810 Moscow, Russia\\
 $^{2}$ Astronomy Department, New Mexico State University, Box 30001,
    Department 4500, Las Cruces, NM 88003-0001, USA}
\date{Received 1995 December 7;
      in original form 1995 July 12}
\begin{document}
\maketitle
\begin{abstract}
 Recently, it was suggested that large concentrations
 or groups of quasars may trace sites of enhanced matter density at
 medium and high redshifts analogous to how galaxy clusters trace them
 in nearby space (Komberg \& Lukash 1994).
 We have checked existing quasar data for the presence of such groups.
 Large Quasar Groups (LQGs) were identified using a well-known cluster
 analysis technique and the following selection criteria:
  (i) LQG must contain at least ten quasars;
  (ii) the number density of quasars in a group must exceed that of the 
       background by at least a factor of two;
  (iii) the majority of quasars in a group must have reliable redshifts.
 Our final list contains 12 such groups, including one reported 
previously. It was found that most of the quasars in these groups come 
from deep homogeneous surveys. Further analysis of the spatial
 distribution of quasars in these surveys has shown that: (i) the 
probability that the detected groups are random is
 rather small (generally a few per cent); (ii) their sizes range from
 $\sim 70$ to $\sim 160h^{-1} Mpc$, which is comparable to sizes of 
nearby rich superclusters; (iii) the detected groups all have redshifts
$0.5<z\leq 2$; (iv) the abundance of the LQGs is comparable with the
abundance of large superclusters at $z\sim 0$, which is consistent with
the idea that quasar groups and superclusters can be evolutionarily 
related. We argue that quasar groups could be a common feature of the 
spatial distribution of medium redshift quasars, and that the quasars in
groups may belong to concentrations of young
 galaxy clusters and groups (distant superclusters) and hence 
 be biased tracers of large-scale structure of matter 
 distribution in the early Universe.
 Theoretical implications as well as other observations 
needed to test this point are discussed.
\end{abstract}
%
\section{Introduction}
%
We have today considerable independent evidence for an existence
of large-scale structures in matter distribution stretching from tens 
of megaparsecs up to a typical scale
$l_{LS}\sim 100-150h^{-1} Mpc$ (unless otherwise stated we assume
$H_0=100h\  km/s/Mpc$, $\Omega_0=1$ and $\Lambda=0$, all scales here 
are comoving). These structures are traced both by galaxies and galaxy
clusters (e.g. Broadhurst et al. 1990; Tully et al. 1992; 
Mo et al. 1992a, 1992b; Einasto \& Gramman 1993; Einasto et al. 1994).
The most important data come from bulk velocities at $z<0.03$
(the Great Attractor), the distribution of galaxies and galaxy clusters 
($z<0.1$), and pencil-beam surveys of galaxies ($z < 0.3$).
These structures develop in the quasi-linear regime of evolution and are 
called great attractors (GAs)/voids:
regions of enhanced/de-enhanced matter density with
$\delta \rho /\rho \sim 10-40\% $ at $l\sim 100h^{-1} Mpc$.
It is now of interest to find how far these large structures extend in 
the past and, thus, when they first appeared in time. The result would 
allow direct dynamical reconstruction of the
primordial density perturbation spectrum in the whole scale range 
$l\in (10,150)h^{-1} Mpc$.

In our first paper (Komberg \& Lukash 1994) we estimated
the perturbation spectrum using quasars. One of the kernels
of the paper was an assumption about recent dynamical epoch when
the first superclusters (the distant GAs observed as LQGs) originated in 
space. Here we see at least two approaches to test this point:
the evolutionary aspects of the quasar correlation function and
a direct search for concentrations or groups of QSOs. We analyzed the 
former test in our second paper
(Komberg, Kravtsov \& Lukash 1994), where indications of  quasar
clustering evolution 
for $z>1$ were found in agreement with other authors
(Kruszewski 1988; Iovino \& Shaver 1988; Iovino, Shaver \& Cristiani 
1991; Mo \& Fang 1993).
In this paper, we report results of a search for and investigation of
large quasar groups.

During the past decade, together with statistical results on quasar 
clustering data, 
evidence has been presented for the existence of four large quasar 
groups. Let us briefly recall their properties:

\begin{enumerate}[(iii)]
  \item The first, found by Webster (1982) at $z\sim 0.37$, consists of 
        four quasars within $\sim 75h^{-1} Mpc$: a close triplet and
         a more distant QSO.
  \item The second LQG, found by Crampton, Cowley \& Hartwick (1987, 
        1989) (hereafter CCH-group) at $z\sim 1.1$, contains 23 quasars 
        within $\sim 60h^{-1} Mpc$ and is thus the richest known group.
  \item The third group of 13 QSOs (Clowes \& Campusano 1991a, 1991b, 
        hereafter CC-group) was found at $z\sim 1.3$, and was reported 
        to have an elongated shape (with long and short sky-projected 
        dimensions of $\sim 150$ and $35h^{-1} Mpc$, respectively) and 
        clumpy inner structure. Graham, Clowes \& Campusano (1995) found
	five additional members of this group and stated the size of
        the group as $150 \times 100 \times 60 h^{-1}Mpc$.
  \item Recently (Graham, Clowes \& Campusano 1995), the Minimal
        Spanning Tree technique 
	was used to search for quasar superstructures in several 
        homogeneous surveys.
        Evidence was found for a new group of quasars at $z\sim 1.9$ and 
        a grouping of Seyfert galaxies at $z\sim 0.19$.
	The former group has a size of 
        $\sim 120 \times 90 \times 20 h^{-1}Mpc$
	and the latter $\sim 60 \times 30 \times 10 h^{-1}Mpc$.
\end{enumerate}
The authors emphasize that both the CCH and CC groups contain 
subclusters of size
$\sim 15-20h^{-1} Mpc$. They also argue that because of a lack of 
further examples the LQGs may be very rare and hence cannot be related 
to the large structures of galaxies which seem to be common features of 
the Universe.
  We discuss this problem below and argue that
if distant quasar groups at $z\geq 1$ trace attractor-like enhancements 
of the total matter density (Komberg \& Lukash 1994), then LQGs must be 
as common as rich superclusters traced by galaxy clusters at $z\leq 0.1$.
We will show as well that, if the latter is true, presently available
wide and deep 
surveys are expected to contain approximately $\sim 3-5$ LQGs each.

Our analysis was carried out as follows. As a first step, we have 
applied  well-known cluster analyses to the largest available quasar 
database -- the V\' eron-Cetty \&  V\' eron (1991) Catalogue of quasars. 
The {\it friend-of-friend} algorithm (e.g. Einasto et al. 1984) was used 
to derive the lists of quasar
cluster candidates for a number of values of clustering radius.
We then excluded clusters containing less than ten quasars and
compared their comoving quasar number density with a background one, 
which was estimated using two deep and wide homogeneous surveys.
Then, we have accepted only clusters with density exceeding the 
background by at least a factor of two. This part of the
investigation is described in Section 2.
Having excluded groups in which most objects had non-reliable redshifts,
we analyzed the properties of the 12 LQGs which were left in the final 
list. Ten of them come from deep homogeneous surveys while two are 
mixtures of quasars from different (though homogeneous) surveys. Our 
further analysis of properties of the quasar groups is based chiefly 
on these surveys: in Section 3 we estimate the probability for a group 
to be a random feature of the
survey and compute an approximate LQG density and their typical size.
Section 4 deals with possible implications of
the obtained results. We summarize our results and conclusions in 
Section 5.
%
\section{The search for Quasar Groups}
%
\subsection{The Catalogue}
Investigation of the spatial distribution of quasars is
strongly hampered by insufficient coverage of the sky by homogeneous 
quasar surveys: they are either deep and narrow or bright and wide. 
The former provide a high enough number density of quasars (the mean 
separation between QSOs in such samples is $\sim 20-50h^{-1}$ Mpc) 
but usually contain a small number of objects. The latter cover large 
areas in the sky and contain many objects but usually provide a very 
low spatial density of quasars (mean quasar separation is 
$\geq 100h^{-1}$ Mpc) which makes them insensitive to the presence of 
large-scale structures. That is why almost all
statistically reliable results on quasar clustering were obtained
with combined samples containing several deep homogeneous surveys (e.g.
Iovino \& Shaver 1988; Andreani \& Cristiani 1992; Mo \& Fang 1993;
Komberg, Kravtsov \& Lukash 1994; Shanks \& Boyle 1994). 
It is clear that for statistical
analysis we need homogeneous data in order to account fairly for all
possible selection effects (though attempts were made to use
inhomogeneous data using the method of {\em normalisation to
the large scales\/} -- see, for example, Shaver 1984 and Kruszewski 
1988). However, for our purpose -- the search for quasar groups, 
we can start with a large inhomogeneous quasar catalog if we consider 
it just like a joint database for known quasars. Let us explain this 
point more clearly. First, we will search for groups similar to those 
discussed earlier (see Section 1),
i.e. we know {\em a priori\/} the kind of structures for which we are 
looking.
Second, available homogeneous quasar surveys differ by methods of 
candidate
selection and may thus miss quasars detected in other surveys in the 
same
area in the sky. Therefore, a catalog containing all  homogeneous 
surveys 
published to date is a good database for the first step of our
investigation. It has, however, a serious shortcoming: the
inhomogeneous nature of the data results in different accuracies of 
quasar coordinates (especially redshifts) -- from reliable to plainly 
wrong. We can overcome this difficulty by excluding ``non-reliable'' 
objects at further stages of our analysis. By so doing we will end up 
with something like a combined quasar sample consisting of several 
homogeneous surveys.

For this purpose we  have used the 5th edition of {\em Catalogue of 
Quasars and Active Nuclei\/} (V\' eron-Cetty \& V\' eron 1991) which 
contains more than 6000 quasars (i.e., the active star-like objects 
with absolute magnitude $M_B<-23$ for $h=0.5$).
The redshifts of some quasars in the Catalogue (marked by an asterisk) 
were estimated from low dispersion slitless spectra and are thus of 
lesser accuracy or even wrong. So, after applying the cluster analysis 
described below, we have discarded from the final list  those quasar 
groups containing a large fraction of QSOs with non-reliable redshifts.
%
\subsection{The search strategy}
%
To find LQGs we have used a cluster analysis method known also as the
{\it friend-of-friend} technique. The kernel of the
method is an objective, automated procedure to separate a set of objects
into individual systems using the following algorithm (see Einasto et al.
1984 for details). Draw a sphere of
radius $R_{cl}$ (the parameter called clustering radius) around each
sample point (in our case, a quasar). If there are other quasars within 
this sphere
they are considered to belong to the same system. 
These quasars are called
{\it friends}. Now draw spheres around these new neighbours and continue 
the procedure using the rule {\em any friend of my friend is my friend}. 
The procedure
stops when there are no more neighbours or {\it friends} to add - a 
system is found. In such systems every object has at least one neighbour 
at distance $l\leq R_{cl}$. It is clear that in this method the choice of
clustering radius $R_{cl}$ is crucial. If  $R_{cl}$ is too
small, then we detect mostly close pairs and triplets. On the other hand,
if $R_{cl}$ is too large all quasars join to form huge systems which have 
a density contrast close to zero. In our case, there is no {\it a priori} 
defined parameter $R_{cl}$ and we have performed the procedure for
a number of values of $R_{cl}$ ranging from $20h^{-1} Mpc$ to 
$60h^{-1} Mpc$ with step $5h^{-1} Mpc$.
These values were chosen for the following reasons: when
$R_{cl}\leq 20h^{-1} Mpc$ we detect systems with less
than five members while at $R_{cl}\geq 60h^{-1} Mpc$ most quasar 
clusters are spread over a large volume and have vanishing density 
contrast.

As a result, we have obtained lists of clusters for each value of 
$R_{cl}$. For each cluster in the list we have calculated its comoving 
dimensions (projected on the
sky $R_{\alpha }$, $R_{\delta }$, and along redshift $R_z$), a rough
estimate of its volume $V_{cl}=R_{\alpha }R_{\delta }R_z$, and the quasar
number density.
We have adopted the following selection criteria to derive LQGs
from the list:
 \begin{enumerate}[(iii)]
  \item a cluster must contain at least ten quasars (i.e., we are going
        to deal with only the largest groups);
  \item the quasar number density in a cluster must exceed that of the
        background by at least a factor of two;
  \item the majority of the quasars in a cluster must have reliable 
        redshifts.
 \end{enumerate}
In fact, these criteria were chosen to obtain LQGs similar to the CCH 
and CC groups. On the other hand, the second criterion makes us sure 
that we deal with real enhancements in quasar number 
density\footnote[1]{A similar density contrast 
criterion was used in the recent paper by Einasto et al. 1994 to 
derive a catalogue of nearby superclusters.} and, certainly, with the 
most prominent ones on
scales $>20h^{-1}$ Mpc. Actually, we look for {\em quasilinear\/} 
structures ($\Delta \rho /\rho <1$)
and if quasars trace the underlying matter distribution with a linear
bias factor $b\sim 3-5$ (typical values for nearby clusters), we should
expect the detection of very prominent
quasar groups somewhere on the level 
$\Delta N/N\simeq b(\Delta \rho /\rho)\sim 1$.
This means that number density in a group exceeds the background 
by a factor of two
\footnote[2]{Our assumption that we deal with {\it quasilinear} 
enhancements in total matter density distribution is strongly backed by 
the huge sizes of the found groups (tipically $\sim 100h^{-1}$ Mpc, 
see below Table 2). For this reason the effect of z-distorsion, 
leading to a general overestimation of density contrast in a collapsing 
system, is negligible in our case.}.
Our group was thus identified when both requirements: a large number of
quasars in the group and a high enough number density for that group, 
were met simultaneously. Thus, we choose a group from among nine lists 
of different clustering radii only when it contains more than ten quasars 
and its density is approximately two times higher than the background 
at a corresponding redshift.
If, for example, we have cluster with more than ten quasars at, say, 
$R_{cl}=35h^{-1}$ Mpc, but with number density contrast still much 
greater than one - we continue searching 
for the same cluster in lists for $R_{cl}>35h^{-1}$ Mpc till 
its density contrast is of order of unity. In some cases,
 $\Delta N/N$ was dropping
rather abruptly in the next $R_{cl}$ list.
When this occurred we took the group from the last list
where $\Delta N/N\geq 1$, which explains why some of identified
 groups have contrast greater than one. 
On the other hand, if we have a cluster of quasars with high density 
contrast but which contains less than 10 quasars for any $R_{cl}$,
it was not included in the list according to the 
first selection criterion.

The background number density of quasars was calculated using  two
large and deep homogeneous surveys (Boyle et al. 1990, Osmer \& 
Hewett 1991). The first survey was derived with UVX technique and is 
thus limited to redshift $z\sim 2.2$. For groups with larger redshifts
we used only
the {\em multicolor} survey by Osmer \& Hewett (1991). 
For all clusters, the faintest member is
within the limiting magnitudes of these surveys. In this case, when 
the background density is 
calculated it is possible to include quasars from surveys
which are fainter than the faintest member of LQG. One can see that the 
density contrast may then be underestimated.
%
\subsection{The Large Quasar Groups found}
%
Having performed the procedure described above we obtained the
final list of 12 quasar groups which fit our selection criteria.

In Table 1 we present only groups in which the majority of quasars have
reliable coordinates and redshifts.
In the first three columns the celestial
coordinates and redshifts of the quasars in the groups are given, followed
by the group extensions in the sky and in redshift directions and
an estimate of their number density contrasts.

It is very interesting to know whether the previously detected groups, 
mentioned in
Section 1, were identified in our analysis. The group discovered by 
Webster (1982) has
been found but does not fit our first criterion.
Five quasars included in the CCH group  (Crampton et al. 1989) appear
as a separate cluster in the list for $R_{cl}=20h^{-1} Mpc$
and group was identified at $R_{cl}=35h^{-1} Mpc$.
For this group $\Delta N/N\sim 1$ and number
of quasars in the cluster is 25 -- 22 of which were included in the 
CCH group (see Section 3 for further details).
The group reported by Clowes \& Campusano (1991) was found only at 
$R_{cl}=60h^{-1} Mpc$ and was not included in the final list as it 
did not fit our second selection criterion (its density contrast is 
less than one). Graham, Clowes \& Campusano 1995 reported detection of
a LQG in the survey by
Osmer \& Hewett 1991 (see Section 1). Unfortunately, the coordinates 
of objects included in the group weren't published. The authors, though,
 present properties
of the group: it consists of 10 quasars, located at $z\sim 1.9$ and has
dimensions $\sim 120\times 90\times 20 h^{-1} Mpc$ (note that as will be
discussed in the next section, the latter dimension most probably is not
real because Osmer \& Hewett survey's geometry is thin in declination 
-- a $\sim 30^{\prime }$ strip which corresponds to 
$\sim 20-30h^{-1} Mpc$ at
$z\sim 1$). We have identified three LQGs in this survey and one of them
(LQG 7, see Tables 1,2) seems to have rather similar properties:
it consists of 10 quasars, is located at $z\sim 1.9$ and has dimensions 
$101\times 92\times 24 h^{-1} Mpc$.

We think that the fact that we have identified the groups found 
in earlier studies shows that the adopted method may be efficient in 
finding quasar groups.
%
\section{Results}
%
 In Section 2 we present a list of  twelve quasar groups.
 Clearly, the use of a heterogeneous catalogue may cause a concern
about the reliability of the results. But we note once more that the 
catalogue
was used only at the very first step of our analysis and was considered
solely as a complete survey of quasar data. In the second step, 
the number density contrasts in groups were estimated using 
{\it homogeneous} surveys.
In fact, we have found that most of the quasars in the detected groups
come from deep and wide homogeneous quasar surveys. This was expected 
because such
surveys provide high quasar densities together with large volumes
surveyed and  hence are most suited for the search of structures 
such as LQGs. In this section we will scrutinize our groups using, 
when possible, the homogeneous surveys from which they originate. 
It gives us the advantage of knowing selection features and geometry 
on the sky of a given survey and holding them under control. On the other 
hand, the background quasar density can be directly estimated using the 
"host" survey where the group is located.
 It allows us to estimate the probability of a
group to appear by chance in such a survey. The most straightforward way
is to generate a random comparison sample with the original selection 
envelope. Random samples were created using the {\em smoothing\/} scheme
(e.g. Mo \& Fang 1993). Redshifts of objects were drawn from a
smoothed version of the redshift distribution while keeping their
celestial coordinates unchanged. The number of quasars in a given bin
of the smoothed redshift distribution (constructed with redshift interval
$\Delta z=0.2$) was calculated by averaging the number of objects in
interval $\Delta z=0.6$ centered at this bin.
In this way, the redshift distribution is randomized and 
selection effects intrinsic to the original survey are preserved. 
Having created the random sample 1000 times and 
performing for each realization the same clustering analysis
(with the value of $R_{cl}$ for which the given group was identified)
 we were able to estimate an empirical probability of a group appearing 
by chance by simply counting
the number of groups found in these random catalogs
 with richness and QSO density 
not less than in the original one and located in the same redshift range.

There is one more thing to note. Several detected groups contain quasars
which come from different surveys. One may be concerned
why observers investigating the same field on the sky may miss quasars 
detected by others. This question is difficult to answer and, obviously, 
in each case individual investigation is required. Our analysis has shown
that, generally, it is a question of different limiting magnitudes,
but there may also be some other explanations: the "alien" quasar lies 
in the sky in the region adjacent to the area of a survey, different 
surveys use different techniques
of candidate selection, exclusion of known quasars, different seeing 
conditions, plate flaws, satellite trails and so on.
%
\subsection{What have we found ?}
%
Here we analyze the LQGs listed in Table 1 
(the groups are arranged according to the
right ascension). For each group we present and briefly
discuss original sources of all the quasars, clustering radius at 
which it was identified, number density of quasars and probability 
for it to be random.
Essential information is summarized in the Table 2.\\
{\bf LQG 1}. This group contains 12 quasars. All redshifts are reliable.
The quasars $0040-3024$, $0040-2919$, $0052-2853$, $0052-2856$ come from
the {\it Large Bright Quasar Survey} (LBQS) (Morris et al. 1991
and references therein). Two
quasars, $0046-2834$ and $0055-2948$, are from Campusano (1991) (the 
first is also present in LBQS, the second is much fainter than LBQS 
limiting magnitude). Quasars $0049-2840$, $0049-2931$, $0052-2847$, 
$0052-2853$, $0057-2835$ come from the deep survey by Boyle et al. 
(1990) (hereafter BFSP).
Two of them were present in the LBQS, the others are too faint. The 
quasar
$0050-2828$ comes from Boyle et al. (1985) and is also present in
Morris et al. (1991), although it is too faint to be included in LBQS. 
The last QSO ($0059-2853$) is from the Warren et al. (1991) survey where 
non-UVX technique was used.

The group was identified at clustering radius $R_{cl}=45h^{-1} Mpc$.
The quasar density in this group is $\sim 3.3 \times 10^{-5}h^3 Mpc^{-3}$.
Clearly, we cannot evaluate the probability to appear by chance for this
group as its objects come from different surveys. However, we note that 
density in this group is approximately 2 times higher than
the density in the BFSP survey (all quasars in the
group are brighter than the limiting magnitude of this survey).\\
{\bf LQG 2}. It contains 12 QSOs, although two of them
($0048-2759$ and $0054-2810$) have redshifts estimated from slitless 
spectra.
As in the previous case, the quasars come from several sources. Three 
QSOs
($0041-2844$, $0046-2914$ and $0048-2901$) are from Morris et al. (1991) 
(LBQS);
$0046-2904$, $0052-2902$ are from Warren et al. (1991). Four objects
($0050-2907$, $0050-2929$, $0052-2830$, $0056-2843$) come from BFSP 
survey but second quasar was also present in the LBQS. The last three 
QSOs ($0048-2804$, $0048-2759$ and $0054-2810$) originate from 
Clowes \& Savage (1983), the quasar selection in this survey was based 
on visual inspection of prism spectra.

The significance of this group cannot be estimated for the same
reasons as for the previous one. The group was identified at 
$R_{cl}=40h^{-1} Mpc$ and its quasar density is $\sim 3.3\times 10^{-5}$ 
(i.e. its density 
contrast is also $\sim 1$).\\
{\bf LQG 3}. This group contains 14 QSOs. All of them, except
$0056-2921$, are from BFSP survey. The QSO $0056-2921$ is from 
Warren et al. (1991). In this survey a multicolor technique was used so 
that it may probably contain quasars missed in UVX BFSP survey.

This group was identified at $R_{cl}=35h^{-1} Mpc$ and its quasar density 
is $\sim 3\times 10^{-5}$. Now we can evaluate the probability for it to 
be random as was described above. The BFSP catalogue contains 34 40-arcmin 
fields, distributed around eight fields. This group resides in the SGP 
field which consists of seven subfields so that some of quasars could 
be easily missed. The estimated probability was found to be rather 
small ($2\% $) so that identification of this group is statistically 
significant.\\
{\bf LQG 4}. There are 14 quasars. Most of them are
also from BFSP but two quasars ($0051-2855$ and $0054-2934$), which 
originate
from Clowes \& Savage (1983), have redshifts estimated from slitless 
spectra. 
Another one ($0050-3001$) is from Savage et al. (1985).

The group was identified at $R_{cl}=35h^{-1} Mpc$ and its density is
$\sim 4.9\times 10^{-5}$, it is 2.5 times higher than quasar
density in the survey at these redshifts. The probability that it appears
by chance
was found to be less than $10^{-3}$ (i.e. no similar groups were found
during 1000 random realizations).\\
{\bf LQG 5}. It consists of 13 QSOs which are all from BFSP
survey (QSF-field). This group was
identified at $R_{cl}=40h^{-1} Mpc$, its density is
$\sim 4\times 10^{-5}$ ($\sim 2$ times higher than the background).  
The quasar
redshift distribution in this field shows a peak at $z\sim 1.7$ which 
is due
to the presence of this group. It may be a random enhancement with 
probability
$\sim 0.03$ (i.e. 30 similar groups were found after 1000 random 
realizations).\\
{\bf LQG 6}. This group consists of 10 QSOs. They all
come from deep homogeneous survey by Osmer \& Hewett (1991) 
(hereafter OH).
It was identified at $R_{cl}=45h^{-1} Mpc$ and its density is
$\sim 7.6\times 10^{-5}$ (it exceeds the background density in OH 
survey by a
factor of 5). The probability for this group to be random is 0.01.\\
{\bf LQG 7}. This group also contains 10 QSOs and 
they all originate from OH survey.
It was identified at $R_{cl}=40h^{-1} Mpc$ and its density is
$\sim 4.6\times 10^{-5}$ (it exceeds the background density in OH 
survey by a
factor of 2). The probability to be random is about 0.19.\\
{\bf LQG 8}. The group consists of the 12 QSOs which are all from 
OH survey.
It was identified at $R_{cl}=45h^{-1}$ Mpc, its density is 
$\sim 5.7\times 10^{-5}$. The probability to be random is about 0.05.\\
{\bf LQG 9}. All quasars are from Crampton, Cowley \&
Hartwick (1989, 1990) and Crampton et al. (1988) (hereafter CFHT survey), 
except one ($1330+2840$)
from Burbidge (1970). The latter is a radio-quiet quasar 
with reliable redshift.
The group was identified at $R_{cl}=35h^{-1} Mpc$, its density is
$2.3\times 10^{-5}h^3 Mpc^{-3}$. The probability to be random is 
$\sim 0.02$.
In the redshift distribution of the CFHT survey there is the excess of
quasars ($\sim 10$ QSOs) at these redshifts.\\
{\bf LQG 10}. This one is the group already known to exist (see CCH).
In our analysis it was identified at $R_{cl}=35h^{-1} Mpc$ and 
consists of
25 quasars (22 of these were included in CCH group). The quasar 
$1340+2843$
is from Wills \& Wills (1979), and two additional QSOs 
($1333+2820$ and $1337+2711$) are from Crampton, 
Cowley \& Hartwick (1990) and Crampton et al.
(1988) respectively.

The number density of quasars in the group is 
$5.1\times 10^{-5}h^3 Mpc^{-3}$.
It is considerably ($\sim 3$ times) higher than the background 
quasar density
in the CFHT survey. There is also peak in the redshift 
distribution at
$z\sim 1.1$.
After 1000 random realizations no similar group was found (i.e. the
probability to be random is less than $10^{-3}$). This is in 
agreement with
CCH who concluded that corresponding probability is vanishly small.\\
{\bf LQG 11}. This group consists of 11 quasars, eight of them
are from the BFSP survey (QSM field), quasar $2200-2019$ is from 
the LBQS,
$2154-1828$ is from Dunlop et al. (1989)
and $2158-1854$ is from Savage et al. 1985.
The group was identified at $R_{cl}=45h^{-1} Mpc$ and its
quasar density is $\sim 2.4\times 10^{-5}h^3 Mpc^{-3}$. The probability 
to be random for the group of 8 quasars in QSM field was found to be 
$\sim 0.05$.
\\
{\bf LQG 12}. This group contains 14 quasars. Twelve of them are from
the BFSP survey, one is from the LBQS ($2200-1958$) and the other 
($2157-2000$)
is from Savage et al. (1976). The latter is a radio-loud quasar and
could, probably, be missed by BFSP as its subfields do not cover the 
whole field. The group
was identified at $R_{cl}=45h^{-1} Mpc$, its density is 
$\sim 3\times 10^{-5}h^3 Mpc^{-3}$. The probability to appear by chance 
was found to be $0.01$.

Finally, we analyze in the same way the CC-group of quasars.
The authors include 13 QSOs in this group.  Though 
Graham, Clowes \& Campusano (1995) found additional members of the group
we use the old number - 10 quasars (three others aren't from that 
survey) -
to estimate probability to be random for this group in the {\em original} 
survey,
because the new extension of the survey was not available to us.
Its sizes are (as computed for all groups in our analysis): 
$R_{\alpha }=134h^{-1} Mpc$,
$R_{\delta }=140h^{-1} Mpc$, $R_z=154h^{-1} Mpc$; quasar number density 
is
$\sim 4.5\times 10^{-6}h^3 Mpc^{-3}$. The estimated probability for such 
structure
to appear by chance in this survey is $\sim 0.03$. The probability is 
small because
number density of quasars in that survey is low (much lower than, say,
in the BFSP survey), and it is unlikely that such structure arises from 
such a low density background. 
 However, one can see that number density of quasars in this group is 
quite low so 
the further observations in
this field are needed which may probably help to find further members of 
this group. 

3D sizes of the groups are difficult to estimate. And one should be very 
careful 
drawing conlusions about LQG sizes. Projected sizes are usually larger 
than 
survey sizes\footnote[1]{
For this reason, the topology of QSO density enhancements is still to be 
clarified, and many of our LQGs are just "objects" determined only in 
z-direction (see also Section 4).}. However, the size in redshift 
direction, 
$R_z$,
should  give us a reasonable estimate of sizes because in that direction 
groups
are not limited.
$R_z$ ranges from $\sim 70$ to $160h^{-1} Mpc$ (see Table 2). 
In case of the OH survey its geometry in the sky,
which is
a narrow strip in declination ($\sim 30^{\prime}$) and wide in right 
ascension
($\sim 12^{\circ}$), 
makes it possible to consider the 
values $R_{\alpha }$ for the groups LQG 6, LQG 7 and LQG 8 found in 
this survey, also as real. 
They are $88$, $101h^{-1}$ and $100h^{-1}$ Mpc respectively, 
i.e. similar
to $R_z$ sizes. Naturally, we can only speculate about the shape of the
groups but there is the real hope that at least in the case of the 
CCH-group
the surveyed area will exceed its size in the near future.
%
\section{Discussion}
%
Now we will briefly consider the implications of our results.
Komberg \& Lukash (1994) suggested that most of the bright QSOs at 
$z \in (1,3)$
form in massive mergers and/or interacting galaxies which occur in 
galaxy protoclusters
and compact groups.
The expected tests are:
\begin{enumerate}[(iii)]
\item  high correlation amplitude for medium-z ($1<z<2$) QSOs with 
correlation 
  radius $\sim 10h^{-1}$ Mpc;
\item high abundance of LQGs at $1\leq z\leq 2$ indicating positions of
 distant presuperclusters which develop later into quasi-linear systems 
like
 the local Great Attractor, Shapley concentration, etc.
\end{enumerate}

Quasars may, therefore, trace enhanced matter density 
regions at medium and high redshifts in the same manner as galaxy groups 
and
clusters do at $z\sim 0$.

The quasar two-point spatial correlation function was estimated recently
by several authors (e.g. Anderson, Kunth \& Sargent 1988; 
Iovino \& Shaver 1988; 
Boyle et al. 1990; Iovino, Shaver \& Cristiani 1991; Andreani \&
Cristiani 1992; Mo \& Fang 1993; Komberg, Kravtsov \& Lukash 1994;
Shanks \& Boyle 1994).
At $z\sim 1-2$ it has an amplitude
 $A_{qq}=\xi (r=1h^{-1}$ Mpc$)=(r_0)^{\gamma }\sim 60-70$ with 
$r_0\sim 10h^{-1}$ Mpc and $\gamma \sim -1.8$ (Komberg, Kravtsov 
\& Lukash 1994). 

How, in the framework of the observational picture described above, 
can we relate QSOs to the underlying density field? 
The answer could be given if we knew 
the mechanism of their formation. Below, we recall some simple 
suggestions 
which can be made to describe the formation of quasars and draw 
relevant conclusions. A necessary condition allowing QSO 
fuelling  in  the early Universe is a high abundance of gas which can be 
captured in the deep potential well provided by a massive 
galaxy. So, from a cosmological point of view, a quasar is just a short 
flash (estimated AGN 
lifetimes are much less than the Hubble time) indicating the position of 
a massive unstable galaxy\footnote[2]{The galaxy (with baryon content 
$M_{b}\geq 10^{11} M_{\odot}$) in a violent 
relaxation period of evolution when a large fraction of the
gravitating gas 
is not rotationally supported and may infall into the galactic
center dissipating radiationally and creating a dense core.
The relation to star 
formation is still uncertain (the gas component is more important 
for QSO burning),
and the host galaxy may even be slightly visible if 
stars had not yet formed in sufficient number.}.  Such a host
galaxy could be young (having just formed from a galactic primordial 
density  
peak) or an older galaxy  which is tidally interacting (or
merging) with another galaxy of similar mass in a group and/or 
protocluster\footnote[3]{Note that the latter QSOs 
(which we call generation-II) may form 
only in galaxies which have lost (at least partially) their
stability as a result of tidal interaction (say, when their intrinsic 
angular
momentum is lowered) and some fraction of matter could infall into the 
center forming the dense bulge which then evolves into an accreting 
gaseous
disk around a black hole.}. The first process (generation-I QSOs) is 
obviously
related with the epoch of galaxy formation, 
whereas the second one (generation-II  
QSOs) proceeds during the formation epoch of compact galaxy groups and 
subclusters\footnote[4]{Recall that our 'compact groups' are 
primordial objects forming in the early Universe (i.e., with small 
dynamical
time) from density perturbation peaks with baryon content $M_{b}\sim
10^{12}-10^{13} M_{\odot}$; nearby observed compact groups are
probably related with loose groups and are just forming at the latest 
epoch of the 
evolution of
loose groups (the latter dynamical time is comparable to the Hubble
time today). Also, we do not discuss here next generations of quasars 
related to other kinds of the environments surviving or originating at 
$z \leq 0.5$: loose groups of galaxies creating small-z 
radio-quiet QSOs distributed like Seyferts, and cD-like galaxies in 
X-ray clusters creating radio-loud QSOs.}. 
The final result of the evolution of primordial compact groups may be 
large elliptical galaxies which seem to reside preferentially in regions 
of enhanced number density of galaxies.

It is worth noting that this picture
of QSO activity originating in merging galaxies is supported by recent 
observations of quasar host galaxies with HST (e.g., McLeod \& Rieke 1995,
Disney et al. 1995, and references therein) which seem to detect 
{\em elliptical} hosts around {\em both radio-loud and radio-quiet} 
quasars.
We think that  at high redshifts ($z\geq 2$)
quasars could form in primordial galaxies (e.g. Haehnelt \& Rees 
1993, Nusser \& Silk 1993) but at medium redshifts ($z\sim 1-2$) when
the galaxy formation rate has decreased the quasar activity could
be driven by mergers in young galaxy clusters and collapsing compact 
groups (Komberg \& Lukash 1994)\footnote[5]{In models with steep 
(CDM-like) spectra of primordial 
density perturbations galaxies form rather late and, therefore, the 
generation-I quasars predominate. However, the generation-II quasars 
become more important in models with extra-power for the perturbations 
on supercluster scale. Certainly, both effects of QSO formation
(primordial and merging galaxies) contribute differently at different 
redshifts, so, the result (QSO spatial distributions,
correlations, etc.) should be very sensitive to the density perturbation 
spectrum.}.
The observational support for this picture comes from high quasar
correlations at $z\sim 1-2$ which are rather similar to those of APM 
galaxy clusters and groups at $z\sim 0$ (e.g. Bahcall \& Chokshi 1993). 
Evidence for the high number density contrast in quasar groups presented 
in this paper 
(see Table 1) also indicates that quasars can be more clustered at 
$z\sim 1-2$ than ordinary galaxies. 

Therefore, we may deal with two different types (generations) of quasars:
\begin{enumerate}[(iii)]
 \item QSOs in primeval massive galaxies; 
 \item QSOs originating in galaxies undergoing tidal 
interactions/mergers 
(with other galaxies) in the first caustics forming in protoclusters and 
       primeval compact groups.
 \end{enumerate}

In this framework, quasar correlations at $z\geq 2$ mirror correlations 
of galaxies at these redshifts, while at $z\sim 1-2$, when most
of quasars may be of the second generation, they trace the underlying 
density field similarly to
the systems they reside in (i.e. the galaxy clusters and 
groups).
It can, for instance, explain the evolution of quasar correlation 
amplitude
(see Iovino \& Shaver 1988; Iovino, Shaver \& Cristiani 1991; 
Mo \& Fang 1993;
Komberg, Kravtsov \& Lukash 1994) from $r_0\sim 2-3h^{-1}$ Mpc 
at $z\geq 2$
up to $r_0\sim 10-13h^{-1}$ Mpc at $z\sim 1-2$ where most of LQGs were 
found.\footnote[2]{We should note, however, that most of the detected 
LQGs
come from catalogs with selection function dropping rapidly at 
$z\geq 2.5$.
Because of that it is yet unclear whether such groups could be present 
at higher redshifts and how much their redshift distribution could 
differ
from the distribution of quasars. So, our finding that LQG abundance 
for a given survey decays faster beyond $z>2$ than the QSO 
abundance, should be tested using the better data in future.}

The LQGs can, therefore, be interpreted as distant superclusters 
analogous
to those traced by galaxy clusters which we observe in the nearby 
Universe
(see Section 1). We will discuss below some observational tests for 
this point,
but first we make a simple consistency estimate comparing 
the abundance of the detected groups of quasars with the abundance of 
superclusters (the groups of clusters) at $z\sim 0$.

The superclusters of galaxy clusters (identified using a similar
clustering procedure) have been recently investigated by Einasto
et al. (1994).  In their list there are eight rich (consisting
of more than 10 clusters) superclusters residing in a volume
$\sim 5.7\times 10^7h^{-3} Mpc^3$. The spatial number density of 
superclusters 
is then $n_{SCL} \sim 1.4\times 10^{-7}h^3 Mpc^{-3}$.
Now, we can estimate the number of LQGs expected in a given survey 
assuming
that the comoving densities coincide $n_{LQG}\sim n_{SCL}$, as follows:
$$N_{LQG}\sim 
\frac {1}{3}{\ }n_{SCL}{\ }\Omega {\ }(r^3(z_{max})-r^3(z_{min})),$$
where $\Omega $ is a surveyed area in steradians; $(z_{min}$, $z_{max})$ 
is the redshift interval of a survey, and $r(z)$ is comoving distance. 
For the BFSP, CCH, and OH surveys the expected number of LQGs is
$N_{LQG}\sim 3, 2, 3$ respectively, while in these surveys we have found
4 (LQG 3, 4, 5, 12), 2 (LQG 9, 10), and 3 (LQG 6, 7, 8) 
groups of quasars.\footnote[1]{We have counted only groups 
which consist of more than ten quasars from the same survey.} 
The agreement shows that the assumption 
$n_{LQG}\sim n_{SCL}$ is indeed plausible.

The idea that quasar groups are basically similar to nearby superclusters
implies that in the regions occupied by LQGs
there are minima of the gravitational potential and thus the LQG regions
are distant great attractors (Komberg \& Lukash 1994). The latter, 
by definition,
are regions of enhanced total density with scales ranging from the
richest cluster size, $l_D\sim 10-15h^{-1} Mpc$ (the largest dynamical
scale of collapsing objects) up to $l_{LS}\sim 100-150h^{-1} Mpc$ 
(the scale of the largest observed structures in galaxy distribution). 
Attractor-like structures 
are generally expanding in the quasi-linear regime, which distinguishes 
them 
markedly from objects collapsing in at least one direction 
(such as galaxy clusters,
filaments, and walls). The idea that groups of clusters
trace mass enhancements is backed by peculiar velocity
measurements: in terms of clusters the local Great Attractor is a modest
concentration (six Abell-like clusters), and similar cluster clumps are 
common 
at larger distances. We can, therefore, conclude that distant attractors 
may also be associated with concentrations of protoclusters 
(presuperclusters) and
hence with large quasar groups. Of course, this hypothesis needs 
further observational verification. Let us briefly discuss some possible 
observational tests.

A search for absorption systems in quasar spectra  
seems  to be one of the most plausible tests for this purpose.
Searching for absorption lines at the redshift of identified LQGs 
lying on the line of sight of background quasars, 
would test the presence of superclusters 
(revealed by possible absorption lines)
at these redshifts. An example of such an investigation is
the analysis of CIV absorption systems in the
spectra of close quasar pair (separation $\sim 18$ arcmin) made by
Jakobsen \& Perryman (1992). Nine additional QSOs were found in the 
same field
and a search for absorption features was carried out. It was suggested 
that
an intervening supercluster may be responsible for the various 
absorption
redshift matches. Further support for the supercluster hypothesis
comes from the fact that both the neighbouring quasars
 and the absorption lines at
the same redshifts appear to lie within a $40-50h^{-1} Mpc$ wide, thick
slab and the geometry of the spatial distribution of both quasars
and absorption systems is rather similar.
So, probing high redshift QSO spectra for possible absorption 
{\em at LQG redshift} may support (or on the contrary exclude) 
the hypothesis that
quasars are associated with superclusters.
Such investigations may provide information about true sizes and
shapes of the underlying superclusters as well.

The imaging of the sky areas in which quasar groups were found
may allow direct measurement of the excess of galaxies as compared to 
control fields. Such investigations were started in the region of the 
CCH
group (Hutchings, Crampton \& Persram 1993; Hutchings, Crampton \&
Johnson 1994).
Deep images of fields around 14 QSOs with narrow-band filters chosen
to detect galaxies near the quasar redshift were obtained (the radius 
of the fields was $\sim 100$ arcseconds, corresponding 
to $\sim 3-5h^{-1}$ Mpc at $z\sim 1$). 
Nine of the 14 QSOs  are from the CCH group (LQG 10). A strong
2-4$\sigma $ excess of faint and blue galaxies associated with CCH QSOs
has been found around seven of them while the other two seem to reside
in sparse groups of galaxies.
We think that this encouraging result should be supported by further 
observations of other fields.

Finally, we would like to emphasize that since the excess of faint blue 
galaxies
in distant ($z \geq 0.5$) galaxy clusters (the Butcher-Oemler effect)
 is related
to the formation and merging of galaxies, we can expect an
excess of faint blue galaxies  in both  proto- and super-clusters 
associated with our LQGs.
%
\section{Conclusions}
%

The main results of the paper are as follows:
\begin{enumerate}[(iii)]
 \item We have identified 12 large quasar groups which meet our
       selection criteria and consist mostly of quasars 
       coming from deep homogeneous surveys.
       The number of quasars in each group is larger than ten and
        the quasar number 
       density excess over the background is larger than a factor of 
       two. 
       The group sizes range from 
       $70$ to $160h^{-1} Mpc$. The quasar group found earlier by 
       Crampton,
       Cowley \& Hartwick (1989) was identified in our analysis as well.
 \item For ten of these groups we have estimated the probability that 
       they are random
       enhancements and found that it is small (usually a few per
       cent, see Table 2). This analysis is based only on
       homogeneous surveys from which the given groups originate.
 \item We thus present further evidence for large-scale structures
       in the quasar distribution of typical size $\sim 100h^{-1} Mpc$. 
 \item We show that the spatial density of the detected quasar groups 
       coincides
       with the number density of nearby superclusters, and argue that
       LQGs may  
       indicate the sites of enhanced total matter density at medium and
       high redshifts (Komberg \& Lukash 1994).
       We thus conclude that quasars at $z\geq 0.5$ may
       trace underlying large-scale structures in matter distribution as
       galaxy clusters and groups do at $z\sim 0$. If this is true,
       more power of the primordial density 
       perturbation field will be required at scale 
        $l_{LS}\sim 100h^{-1} Mpc$ (the 'blue bump')
       than is
       usually assumed from  galaxy and cluster distributions at 
       $z\sim 0$.
\end{enumerate}
To summarize, we can say that the present evidence for large-scale
structures extending over the wide range of redshifts  $z\sim 0.5-2$ 
and the hypothesis that quasars in groups are 
associated with young galaxy clusters,
although far from being conclusive, allows for various verifications 
via both theory and observations and may fuel further detailed 
investigations on this subject.
%
\section{Aknowledgments}
%

We would like to thank the staff of the theoretical division of the 
Astro Space Center for productive discussions; Angela Iovino 
for suggesting the 
way to estimate empirical probability for a group to be random and for
valuable discussions; the referee, David Crampton, for useful comments; 
Neal Miller for the help in improving the presentation of the paper;
and Elena Mikheeva for the help in preparation of the manuscript.
This work was partly supported by Russian Foundation
for Fundamental Research (93-02-2929), 
International Science Foundation (MEZ300), and COSMION 
(cosmomicrophysics). V.N.L. would also like to thank the German Science 
Foundation for financial support during a one-month-stay in AIP (Potsdam) 
where the paper has been completed.


\pagebreak

\begin{table*}
\begin{minipage}{115mm}
\caption[ ]{ Large quasar groups found in our search. 
Coordinates of quasars are given as they are in the catalogue. 
The parameters are 
comoving sizes of a group in megaparsecs and approximate quasar 
number density contrast
(see Section 3 for details).  }
\begin{flushleft}
\begin{tabular}{cccccccc} \hline
            &               &          &            
         &            &               &          &               \\
  R.A.      & Decl.         & Redshift & Parameters          
&  R.A.      & Decl.         & Redshift & Parameters          \\         
            &               &          &                     
&            &               &          &                \\ \hline \\
            &    LQG 1      &          &                     
&            &     LQG 4     &          &                     \\
00 40 33.2  &   -30  24 08  &   0.609  & $R_{\alpha }= 102  $
&  00 49 37.9&   -29 08 38   &    1.855 &    $R_{\alpha }= 46$\\
00 40 46.3  &   -29  19 40  &   0.624  & $R_{\delta }= 42  $ 
&  00 49 42.1&   -29 34 08   &    1.868 &    $R_{\delta }= 60$\\ 
00 46 18.0  &   -28  34 01  &   0.632  & $R_z= 96$           
&  00 49 46.2&   -29 21 40   &    1.856 &    $R_z= 104$       \\ 
00 49 42.3  &   -28  40 27  &   0.639  & $\Delta N/N \sim 1$ 
&  00 49 47.8&   -29 35 26   &    1.920 &    $\Delta N/N\sim 1$   \\ 
00 52 18.0  &   -28  47 34  &   0.639  &                     
&  00 50 28.6&   -30 01 01   &    1.922 &                       \\ 
00 52 51.4  &   -28  53 24  &   0.634  &                     
&  00 51 17.6&   -28 55 10   &    1.94  &                       \\ 
00 55 38.5  &   -28  28 23  &   0.648  &                     
&  00 50 12.9&   -29 07 30   &    1.976 &                       \\ 
00 57 01.9  &   -28  35 42  &   0.662  &                     
&  00 51 51.4&   -29 18 34   &    1.987 &                       \\ 
00 59 10.6  &   -28  53 37  &   0.62  &                      
&  00 53 18.2&   -28 40 23   &    1.964 &                       \\ 
00 55 43.3  &   -29  48 59  &   0.668  &                     
&  00 53 30.9&   -28 36 26   &    1.920 &                       \\ 
00 52 40.9  &   -28  56 48  &   0.602  &                     
&  00 53 37.0&   -28 43 11   &    1.933 &                       \\ 
00 49 27.5  &   -29  31 18  &   0.601  &                     
&  00 53 19.5&   -29 21 51   &    2.029 &                       \\ 
            &    LQG 2      &          &                     
&  00 53 52.5&   -29 30 14   &    1.969 &                       \\ 
00 41 24.2  &   -28  44 06  &   0.839  &  $R_{\alpha }= 84$
&  00 54 33.1&   -29 34 04   &    2.01  &                       \\ 
00 46  2.4  &   -29  04 53  &   0.84  &  $R_{\delta }= 39$ 
&             &    LQG 5     &          &                       \\
00 48 47.0  &   -28  04 19  &   0.840  &  $R_z= 111$       
&  03 35 52.7 &  -44 06 54    &    1.679 & $R_{\alpha }= 45$  \\          
00 50 28.2  &   -29  07 42  &   0.852  &  $\Delta N/N \sim 1$   
&  03 37 37.6 &  -44 21 29    &    1.661 & $R_{\delta }= 49$  \\ 
00 50 36.9  &   -29  29 13  &   0.830  &                     
&  03 37 06.4 &  -44 11 54    &    1.609 & $R_z= 146$         \\ 
00 48 22.5  &   -27  59 40  &   0.87  &                      
&  03 41 32.7 &  -44 58 14    &    1.662 & $\Delta N/N\sim 1$     \\ 
00 52 44.5  &   -29  02 31  &   0.84  &                     
&  03 41 50.7 &  -45 17 41    &    1.615 &                      \\ 
00 56 12.5  &   -28  43 26  &   0.828  &                     
&  03 42 13.5 &  -45 03 04    &    1.700 &                      \\ 
00 54 22.5  &   -28  10 27  &   0.80  &                      
&  03 39 48.1 &  -44 52 00    &    1.745 &                      \\ 
00 52 33.8  &   -28  30 45  &   0.779  &                     
&  03 38 06.1 &  -44 18 32    &    1.762 &                      \\ 
00 46 50.7  &   -29  14 40  &   0.781  &                     
&  03 38 06.1 &  -44 22 12    &    1.733 &                      \\ 
00 48 24.2  &   -29  01 46  &   0.783  &                     
&  03 39 22.9 &  -44 16 54    &    1.764 &                      \\ 
            &    LQG 3      &          &                     
&  03 40 12.0 &  -44 03 38    &    1.751 &                      \\ 
00 48 08.1  &  -28  18 45   &  1.322   & $R_{\alpha }= 67$ 
&  03 40 15.8 &  -44 26 08    &    1.792 &                      \\ 
00 50 24.1  &  -28  17 39   &  1.331   & $R_{\delta }= 57$ 
&  03 42 07.2 &  -44 56 18    &    1.827 &                      \\ 
00 50 28.3  &  -27  47 29   &  1.355   & $R_z= 123$        
&             &     LQG 6     &          &                      \\ 
00 51 39.5  &  -28  46 47   &  1.338   & $\Delta N/N\sim 1$    
&  12 07 34.1 &  -11 00 32    &    1.555 & $R_{\alpha }= 88$  \\           
00 53 33.8  &   -28  36 49  &   1.306  &                     
&  12 07 44.8 &  -11 01 17    &    1.592 & $R_{\delta }= 16$  \\
00 53 15.2  &   -29  24 41  &   1.331  &                     
&  12 08 40.9 &  -11 10 16    &    1.571 & $R_z= 94$          \\ 
00 53 35.0  &   -29  22 47  &   1.303  &                     
&  12 12 11.5 &  -10 56 33    &    1.626 & $\Delta N/N \sim 4$     \\ 
00 55 42.9  &   -28  50 11  &   1.276  &                     
&  12 12 59.1 &  -10 50 34    &    1.590 &                      \\ 
00 56 34.5  &   -29  05 30  &   1.341  &                     
&  12 14 37.1 &  -10 45 10    &    1.535 &                      \\ 
00 56 51.8  &   -29  01 46  &   1.281  &                     
&  12 15 02.1 &  -11 03 38    &    1.498 &                      \\ 
00 56 20.8  &   -29  15 34  &   1.255  &                     
&  12 16 23.1 &  -11 00 17    &    1.551 &                      \\ 
00 55 15.2  &   -28  39 22  &   1.366  &                     
&  12 16 38.6 &  -11 04 51    &    1.538 &                      \\ 
00 55 19.1  &   -28  52 38  &  1.388   &                     
&  12 16 50.9 &  -10 48 07    &    1.505 &                      \\ 
00 56 50.7  &   -29  21 49  &  1.40   &                     
&              &               &          &                      \\   
            &               &          &                     
&            &               &          &                        \\ 
            &               &          &                     
&            &               &          &                        \\
\hline
\end{tabular}
\end{flushleft}
\end{minipage}
\end{table*}

\pagebreak

\setcounter{table} 0

\begin{table*}
\begin{minipage}{115mm}
\caption[ ]{ {\it continued} }
\begin{flushleft}
\begin{tabular}{cccccccc} \hline
            &               &          &                               
&             &               &          &                     \\
  R.A.      & Decl.         & Redshift & Parameters                    
&   R.A.      & Decl.         & Redshift& Parameters          \\
            &               &          &                               
&             &               &          &          \\
 \hline \\
            &     LQG 7     &          &                               
&             &               &          &                       \\
12 23 31.3  &   -10  51 13  &   1.872  & $R_{\alpha }= 101$            
& 13 36 50.7  &    28  20 38  &   1.113  &                      \\    
12 23 31.9  &   -11  13 41  &   1.828  & $R_{\delta }= 24$             
& 13 36 47.8  &    28  23 41  &   1.124  &                      \\    
12 24 55.8  &   -11  04 16  &   1.883  & $R_z= 92$                     
& 13 37 26.7  &    27  26 11  &   1.120  &                      \\    
12 26 37.0  &   -11  05 56  &   1.900  & $\Delta N/N \sim 1$              
& 13 38 01.2  &    27  35 50  &   1.140  &                      \\    
12 27 36.0  &   -10  51 56  &   1.960  &                               
& 13 38 21.4  &    27  36 00  &   1.139  &                      \\    
12 29 25.8  &   -10  52 42  &   1.904  &                               
& 13 35 13.3  &    26  51 08  &   1.096  &                      \\    
12 24 49.6  &   -11  16 59  &   1.979  &                               
& 13 39 25.2  &    27  33 24  &   1.095  &                      \\    
12 30 23.7  &   -10  43 25  &   1.934  &                               
& 13 38 41.4  &    27  40 26  &   1.175  &                      \\    
12 33 13.7  &   -10  50 27  &   1.902  &                               
& 13 40 40.6  &    28  02 27  &   1.164  &                      \\    
12 33 01.6  &   -10  57 40  &   1.884  &                               
&  13 33 59.9  &    27  02 54  &   1.068  &                      \\          
&  LQG 8      &          &                               &  13 36 10.8  
&    26  53 38  &   1.088  &                  \\   
12  16   1.6&   -10 47 43   &   2.119  & $R_{\alpha }= 100$            
&  13 37 50.1  &    27  11 41  &   1.205  &                     \\
12  17  59.2&   -10 55 23   &   2.092  & $R_{\delta }= 20$             
&  13 39 31.6  &    27  22 57  &   1.175  &                     \\
12  19  56.4&   -11 13 29   &   2.105  & $R_z= 104$                    
&  13 39 49.5  &    27  24 35  &   1.185  &                     \\
12  22   6.3&   -11 06 49   &   2.171  & $\Delta N/N \sim 2$              
&  13 37 05.8  &    27  38 47  &   1.056  &                     \\
12  19  31.2&   -11 13 35   &   2.194  &                               
&  13 36 57.5  &    27  43 27  &   1.047  &                     \\
12  23  32.2&   -10 51 24   &   2.191  &                               
&  13 39 47.0  &    27  56 45  &   1.036  &                     \\
12  24  35.4&   -10 56 47   &   2.142  &                               
&  13 39 59.5  &    26  58 06  &   1.053  &                     \\
12  18   0.6&   -11 02 11   &   2.192  &                               
&  13 40 36.4  &    28  43 10  &   1.037  &                     \\
12  18  35.4&   -10 48 40   &   2.241  &                               
&              &      LQG 11   &          &                     \\
12  25   3.0&   -10 48 15   &   2.242  &                               
&  21 54 12.1  &   -18  28 04  &   0.668  &  $R_{\alpha }= 63$\\12  16  14.0
&   -10 46 54   &   2.275  &                               &  21 58 00.0  
&   -18  54 00  &   0.7  &  $R_{\delta }= 42$ \\
12  24  35.2&   -11 07 07   &   2.290  &                               
&  21 58 14.2  &   -18  55 48  &    0.687  &  $R_z= 157$      \\
            &      LQG 9    &          &                               
&  22 00 32.1  &   -20  19 52  &    0.671  &  $\Delta N/N\sim 2$  \\
13 32 23.9  &    27  34 14  &   1.866  & $R_{\alpha }= 88$             
&  21 56 15.8  &   -19  29 36  &    0.725  &                    \\
13 33 34.8  &    28  08 36  &   1.886  & $R_{\delta }= 134$            
&  21 59 30.7  &   -19  31 46  &    0.728  &                    \\
13 33 54.2  &    28  40 16  &   1.908  & $R_z= 66$                     
&  22 06 14.8  &   -20  19 03  &    0.682  &                    \\
13 34 54.4  &    27  28 01  &   1.909  & $\Delta N/N\sim 1$              
&  22 03 30.1  &   -19  20 01  &    0.649  &                    \\
13 35 19.0  &    28  29 02  &   1.865  &                               
&  22 01 09.1  &   -18  57 08  &    0.615  &                    \\
13 34 03.2  &    27  22 32  &   1.931  &                               
&  22 03 01.7  &   -18  46 26  &    0.626  &                    \\
13 34 30.4  &    27  36 04  &   1.95   &                               
&  22 03 25.8  &   -18  50 17  &    0.619  &                    \\
13 35 24.8  &    27  16 46  &   1.928  &                               
&              &      LQG 12   &           &                    \\
13 36 38.4  &    27  27 03  &   1.922  &                               
& 21 57 16.6  &   -19  40 10  &   1.206  & $R_{\alpha }= 80$    \\
13 33 39.7  &    26  43 45  &   1.936  &                               
& 21 57 21.8  &   -20  00 11  &   1.198  & $R_{\delta }= 46$    \\
13 33 29.9  &    26  17 59  &   1.899  &                               
& 21 58 29.9  &   -19  03 01  &   1.240  & $R_z= 155$           \\
13 33 42.0  &    26  17 47  &   1.926  &                               
& 21 59 23.9  &   -19  26 17  &   1.170  & $\Delta N/N\sim 4$     \\
13 34 41.7  &    26  14 33  &   1.876  &                               
& 22 00 35.6  &   -19  29 37  &   1.165  &                      \\
13 36 17.2  &    25  31 56  &   1.88   &                               
& 22 00 50.5  &   -19  49 40  &   1.168  &                      \\
13 36 36.5  &    26  48 58  &   1.868  &                               
& 22 00 28.8  &   -19  52 24  &   1.277  &                      \\
13 38 17.6  &    25  51 28  &   1.877  &                               
& 22 00 44.5  &   -18  49 31  &   1.288  &                      \\
13 38 23.3  &    26  37 05  &   1.841  &                               
& 22 00 54.5  &   -19  58 21  &   1.260  &                      \\
13 41 30.9  &    25  39 47  &   1.896  &                               
& 21 57 46.5  &   -19  34 51  &   1.142  &                      \\
            &      LQG 10   &          &                               
& 21 59 40.8  &   -18  50 10  &   1.155  &                      \\
13 33 20.3  &    28  20 19  &   1.095  &$R_{\alpha }= 51$              
& 22 03 15.6     &   -18  36 30  & 1.179&                      \\
13 33 43.1  &    27  43 04  &   1.116  &$R_{\delta }= 59$        
& 22 01 59.6  &   -19  05 29  &   1.123  &                      \\
                                                                  
13 35 06.7  &    27  51 53  &   1.121  &$R_z= 164$                     
& 22 05 59.3  &   -19  40 14  &   1.280  &                      \\ 
                                                                 
13 35 20.1  &    28  19 59  &   1.124  &$\Delta N/N\sim 2$             
&             &               &          &                      \\
                                                                  
13 35 48.4  &    28  20 23  &   1.086  &                               
&             &               &          &                      \\
13 36 10.5  &    28  18 10  &   1.116  &                               
&             &               &          &                      \\    
            &               &          &                               
&             &               &          &                       \\ 
\hline          
\end{tabular}   
\end{flushleft}
\end{minipage} 
\end{table*}     

\begin{table*}
\begin{minipage}{115mm}
  \caption{Summary of LQG analysis.}
  \begin{tabular}{@{}cccccc}
    Group         &  Number   &  z    & $R_z$,    &  Density,
                         &   Probability    \\
                  &   of QSOs &       & $h^{-1} Mpc$ &  
$\times 10^{-5}h^3 Mpc^{-3}$   &   to be random   \\
    LQG 1         &     12    &  0.6  &    96       
&      3.3                        &      --         \\
    LQG 2         &     12    &  0.8  &    111      
&      3.3                        &      --         \\
    LQG 3         &     14    &  1.3  &    123      
&      3.0                        &      0.02        \\
    LQG 4         &     14    &  1.9  &    104      
&      4.9                        &      $< 10^{-3}$ \\
    LQG 5         &     13    &  1.7  &    146      
&      4.0                        &      0.03        \\
    LQG 6         &     10    &  1.5  &    94       
&      7.6                        &      0.01        \\
    LQG 7         &     10    &  1.9  &    92       
&      4.6                        &      0.19        \\
    LQG 8         &     12    &  2.1  &    104      
&      5.7                        &      0.05        \\
    LQG 9         &     18    &  1.9  &    66       
&      2.3                        &      0.02 \\
    LQG 10        &     25    &  1.1  &    164      
&      5.1                        &      $< 10^{-3}$        \\
    LQG 11        &     11    &  0.7  &    157      
&      2.4                        &      0.05        \\
    LQG 12        &     14    &  1.2  &    155      
&      3.0                        &      0.01        \\
CC-group          &     13    &  1.3  &    154      
&      0.5                        &      0.03        \\
\end{tabular}
\end{minipage}
\end{table*}

\end{document}